\documentclass[nofootinbib,amsfonts,preprint,aps]{revtex4}
\usepackage{graphicx}
\begin{document}

\title{Emergence of string-like physics from Lorentz invariance in
  loop quantum gravity\footnote{Essay written for the Gravity Research
  Foundation 2014 Awards for Essays on Gravitation}}

\author{Rodolfo Gambini$^{1}$,
Jorge Pullin$^{2}$}
\affiliation {
1. Instituto de F\'{\i}sica, Facultad de Ciencias, 
Igu\'a 4225, esq. Mataojo, 11400 Montevideo, Uruguay. \\Email: rgambini@fisica.edu.uy \\
2. Department of Physics and Astronomy, Louisiana State University,
Baton Rouge, LA 70803-4001.\\ Email: pullin@lsu.edu\\ \\
{Submitted March 29th. 2014}}

\begin{abstract}
  We consider a quantum field theory on a spherically symmetric
  quantum space time described by loop quantum gravity. The spin
  network description of space time in such a theory leads to
  equations for the quantum field that are discrete. We show that to
  avoid significant violations of Lorentz invariance one needs to
  consider specific non-local interactions in the quantum field theory
  similar to those that appear in string theory. This is the first
  sign that loop quantum gravity places restrictions on the type of
  matter considered, and points to a connection with string theory
  physics.
\end{abstract}

\maketitle

There has been recent progress in studying quantum field theory on a
quantum space-time in loop quantum gravity \cite{qftqst}. In
particular spherically symmetric quantum space-times were exactly
solved and can be used as a background to study quantum matter fields
living on black hole space times\cite{spherical}. If one considers
background quantum states that are peaked around a definite value of
the ADM mass, the main effect of the quantum nature of the background
is that the equations that result for the field theory are discrete,
as if it had been placed on a lattice \cite{hawking}. Except that in
this case the lattice is not a computational tool, but is fundamental
in nature, it is given by the discreteness of space-time in loop
quantum gravity. In particular, in spherical symmetry one has that the
areas of the spheres of symmetry are given by an integer multiple of
the Planck area. The presence of a lattice structure leads to
violations of Lorentz invariance. How to recover the approximate
continuum behavior of the quantum field theory in regions of small
curvature, for low energies and for interacting quantum field theories
has to be addressed. It is well known that although the violations
appear at Planck scale only, Collins {\em et al.}  \cite{collins} have
argued that as soon as one considers interacting quantum field
theories and certain diagrams requiring renormalization, the
violations become large and experimentally unacceptable. In this essay
we would like to argue that to avoid such problems, one needs to
consider interacting quantum field theories with non-local
interactions, of a kind similar to those that appear in string field
theory \cite{woodard}.

As we shall see, some degree of non-locality is expected since the
background quantum space-time will generically be in a state that is
the superposition of different values of the ADM mass. This is
tantamount to consider superpositions of lattices, which leads to
spatial non-locality. To preserve Lorentz invariance, however, one is
led to consider non-locality in space-time as those considered in
string theory.

Let us briefly recall our previous quantum treatment of vacuum
spherically symmetric space-times \cite{spherical}. The quantum states
that solve the diffeomorphism and Hamiltonian constraints are based on
one dimensional spin networks. A basis is labeled by a graph $g$, a
vector of valences $\vec{k}$ and the value of the ADM mass, $\vert
\tilde{g},\vec{k},M\rangle$, where $\tilde{g}$ is the equivalence
class of graphs under diffeomorphisms of $g$ . The integer values of
the vector of valences characterize the value of the area of the
spherical surfaces of symmetry in terms of Planck units. On these
states one can realize the Hamiltonian of a scalar matter (test) field
as a Dirac observable acting on the vacuum gravitational states
\cite{hawking}. The expectation value of the Hamiltonian taken on the
vacuum gravitational states can be used to construct the equations for
the matter field. In the particular case in which the vacuum quantum state is
peaked in the ADM mass, the resulting equations for matter look like
discrete versions of the continuum equations for the field. The
discreteness is induced by the background quantum geometry.

Generically, the background quantum state will not be in an eigenstate
of the ADM mass. A spread of the ADM mass translates itself in a
fuzziness of the resulting discrete quantum field theory. To see this,
we notice that the matter field lives at the vertices of the spin
network of the background quantum state. It is therefore characterized
by a vector of values $\vec{\phi}$ representing the values of the
field at the vertices of the spin network. The complete quantum state for
gravity and matter is given by $\vert \tilde{g}, \vec{k}, M,
\vec{\phi}\rangle$. To make contact with the traditional picture of
quantum field theory on a classical space-time, one needs to make
several assumptions \cite{spherical,hawking}, and in particular to
identify the components of the vector $\vec{\phi}$ with values of the
fields at particular coordinate points $\phi(r)$. Such identification
will generically depend on the complete quantum state of matter and
gravity. If such a state is in a superposition of values of the mass
(and also generically, of the valences $\vec{k}$) this will imply that
each value $\phi(r)$ will correspond to a superposition of components
of $\vec{\phi}$. For instance, if we consider a Gaussian spread in the
mass, this will imply a Gaussian superposition in the fields of
different points in the spin network, which in Fourier space
translates into the fields picking up a factor $\exp(-\sigma(\Delta M)
\left(\vec{p}\,\right)^2)$ with $\sigma(\Delta M)$ a function dependent on the width
of the superposition in the mass, which we take to be of Planck scale
$\Delta M\sim \ell_{\rm Planck}$. To derive the factor in three
dimensions one needs to expand the plane waves one uses to compute the
Feynman propagators in terms of spherical waves. One can show that the
fuzziness in the radial direction induces one in the angular
directions. 

Such factors can help with the violations of Lorentz invariance due to
the presence of the discreteness, although they do not solve them
completely. One will need an additional ingredient.  To see this, let
us start by briefly recalling the argument of Collins {\em et al.}
\cite{collins}. They consider a model consisting of a Yukawa coupling
of a scalar field and a fermion, but the essence of the argument
applies to generic interacting quantum field theories. They study
diagrams that involve integrals over internal momentum lines. Such
integrals diverge and have to be renormalized. They focus on a
particular quantity, the self-energy and compute its second derivative
with respect to $p_0$ and with respect one of the spatial components
of the momentum, for instance $p_1$ and subtract them, evaluating them in
$p^\mu=0$. If the quantity is Lorentz invariant (i.e. function of $p^\mu p_\mu$), such subtraction
should vanish. Computing the self energy with the usual propagators in
the continuum the result is indeed zero with a suitable regularization
procedure that respects Lorentz invariance. However, they show that if
one considers propagators that violate Lorentz invariance, even if the
violation only happens at very high energies, the result is
non-vanishing and in fact it is bounded below by finite quantity
that would lead to incompatibilities with experimental evidence.

If one were to repeat Collins {\em et al.}'s calculation for the
quantum field theory on a quantum space time we are considering, the
fact that effectively the resulting theory is discrete implies that
dispersion relations for the fields are those of a lattice theory and
therefore not Lorentz invariant. A straightforward calculation
\cite{polchinski}  shows that the second derivatives of the
self energy do not cancel and one has large violations of Lorentz
invariance. What about the extra exponential factors in the momentum
of the field we noticed?  Those factors apparently would help since
they tend to cutoff large values of the momentum and therefore confine
the calculation to a regime where the dispersion relation of the
lattice theory is approximately Lorentz invariant. However, by
imposing a cutoff only in the spatial part of the momentum, one is
further violating Lorentz invariance (the integral in $p^0$ is
unbounded while the other is damped by the Gaussian factor) and one
can show that the second derivatives do not cancel, again implying
unacceptably large violations of Lorentz invariance. This was actually
explicitly studied by Collins {\em et al.} \cite{collins}.

As we stated, we need an additional ingredient to restore Lorentz
invariance at low energies. This additional ingredient is to consider
a theory with a non-local interaction that will create a similar bound
as the one we had in spatial momenta in the time component of the
momentum. For simplicity, let us consider a $\lambda \phi^4/4!$
theory. We replace the interaction in momentum space by,
\begin{equation}
  \frac{\lambda}{4!} 
\left[\exp\left(-\alpha^2 \left(p_0^2-\left(\vec{p}\,\right)^2\right)^2\right)\phi(p_0,\vec{p}\,)\right]^4,
\end{equation}
where $\alpha$ is a function of $\Delta M$. This interaction would
correspond in position space to a non-local interaction. There are two
requirements guiding us: the interaction should be Lorentz invariant
and the exponential nature of the factor is to make it compatible with
the non-locality discussed before. This does not determine the
interaction uniquely, but significantly constrains its functional
form. The extra factors present imply that the integrals in momentum
space are confined to regions close to the light cone. Combined with
the Gaussian factors in spatial momentum we discussed before, they
imply that both the integrals on the spatial and temporal components
of the momentum are confined to regions where the dispersion relations
of the discrete treatment are approximately Lorentz invariant. One can
show that the calculation of the second derivatives of the self energy
considered by Collins {\em et al.}  yields a result that can be made
arbitrarily small by choosing appropriate functions of the mass width
$\alpha$ and $\sigma$ (see appendix). Given that the lattice
separations imposed by the quantization of area in loop quantum
gravity are very small, Lorentz invariance holds very accurately up to
energies considerably higher than grand unification energies. The
resulting effective theory one recovers is the usual $\lambda
\phi^4/4!$ theory.

The non-local exponential types of interactions we are considering
have been extensively studied in the context of string theories
\cite{woodard}. In general non-local theories have problems related to
ghosts and the Ostrogradsky instability. However, there exist more
benign non-local theories that arise as effective field theories in
string theory and the type of theory we are considering here is of
that class. It should be noted that this is the first instance in
which loop quantum gravity imposes restrictions on the matter content
of the theory. Up to now loop quantum gravity, in contrast to
supergravity or string theory, did not appear to impose any
restrictions on matter.  Here we are seeing that in order to be
consistent with Lorentz invariance at small energies, limitations on
the types of interactions that can be considered arise. The
limitations include having to involve constants that are determined by
properties of the geometry. One can also expect that the values of the
coupling constants will relate to their bare values through expressions
involving properties of the geometry as well, like fluctuations in the
Schwarzschild radius.

Summarizing, we have argued that in order to have Lorentz invariance
at low energies in quantum field theories on the quantum space-times
that arise in loop quantum gravity one needs to start with matter
fields that have interactions of the type that arise in effective
theories stemming from string theory. This implies a significant
restriction on kind of interactions in the matter content of the
theory and opens the possibility for contacts between the physics of
loop quantum gravity and string theory.

We wish to thank Abhay Ashtekar for comments.
This work was supported in part by grant NSF-PHY-1305000, funds of the
Hearne Institute for Theoretical Physics, CCT-LSU and Pedeciba.

\section{Appendix: A model calculation}

The calculation of the second derivatives of the self energy for a
discrete theory with the weighting functions we consider is
lenghty. To illustrate its behavior we exhibit here a simple example
that captures what is going on. Consider the following integral in two
dimensions,
\begin{equation}
  I = \int_{-\infty}^\infty dx dy \frac{x^2-y^2}{\left(1+x^2+\frac{y^2}{1+\sigma y^2}\right)^2}.
\end{equation}
If $\sigma=0$ the denominator is invariant under rotations and the
integral vanishes. This would be the analogue of the difference of the
derivatives of the self-energy diagram calculation in the continuum
when Lorentz invariance holds. The case $\sigma$ different from zero
would correspond to the calculation on the lattice, where Lorentz
invariance is broken, in this example represented by the breakage of
rotational invariance. $\sigma$ would play the role of lattice
spacing. When it is not zero, the different behavior in $x$ and $y$
for large values implies the integral does not vanish by a significant
amount.  However, if one considers the integral with weighting factors
similar to the ones considered in this essay,
\begin{equation}
  I = \int_{-\infty}^\infty dx dy \frac{x^2-y^2}{\left(1+x^2+\frac{y^2}{1+\sigma y^2}\right)^2} e^{-\alpha
    x^2} e^{-\beta\left(x^2+y^2\right)}
\end{equation}
with the first weighing factor representing the one stemming from
fuzziness and the second representing the non-local interaction (in
this case rotationally invariant, in the case of interest, Lorentz
invariant), one can see that with suitable choices of $\alpha$ and
$\beta$ one can make the integral as small as desired. This is due to
the exponentials cutting off the integral in the region (large values of
$x,y$) that yielded the non-vanishing contributions. More precisely
one needs $\alpha\sim \beta$ and both larger than $\sigma$.

\end{document}